\documentclass[twocolumn,showpacs,preprintnumbers,amsmath,amssymb,superscriptaddress]{revtex4-1}

\usepackage{graphicx}
\usepackage{dcolumn}
\usepackage{bm}
\usepackage{ulem}
\usepackage{pifont}
\usepackage{natbib}
\usepackage{titlesec}
\usepackage{soul}
\usepackage{hyperref}
\hypersetup{
colorlinks = true,
urlcolor   = blue,
linkcolor  = blue,
citecolor  = blue
}

\begin{document}

\preprint{APS/PRB}
\title{Intertwined Magnetic Sub-Lattices in the Double Perovskite Compound LaSrNiReO$_6$}

\author{Ola~Kenji~Forslund}
   \email{okfo@kth.se}
\affiliation{Department of Applied Physics, KTH Royal Institute of Technology, SE-106 91 Stockholm, Sweden}
\author{Konstantinos~Papadopoulos}
\affiliation{Department of Physics, Chalmers University of Technology, SE-41296 G\"oteborg, Sweden}
\author{Elisabetta~Nocerino}
\affiliation{Department of Applied Physics, KTH Royal Institute of Technology, SE-106 91 Stockholm, Sweden}
\author{Gerald~Morris}
\author{Bassam~Hitti}
\author{Donald~Arseneau}
\affiliation{TRIUMF, 4004 Wesbrook Mall, Vancouver, BC, V6T 2A3, Canada}
\author{Vladimir~Pomjakushin}
\affiliation{Laboratory for Neutron Scattering and Imaging, Paul Scherrer Institute, CH-5232, Villigen, PSI, Switzerland}
\author{Nami~Matsubara}
\affiliation{Department of Applied Physics, KTH Royal Institute of Technology, SE-106 91 Stockholm, Sweden}
\author{Jean-Christophe~Orain}
\affiliation{Laboratory for Muon Spin Spectroscopy, Paul Scherrer Institute, CH-5232 Villigen PSI, Switzerland}
\author{Peter~Svedlindh}
\affiliation{Department of Materials Science and Engineering, Uppsala University, Box 35, SE-751 03 Uppsala, Sweden}
\author{Daniel~Andreica}
\affiliation{Faculty of Physics, Babes-Bolyai University, 400084 Cluj-Napoca, Romania}
\author{Somnath~Jana}
\affiliation{Centre for Advanced Materials, Indian Association for the Cultivation of Science, Jadavpur, Kolkata, 700032, India}
\author{Jun~Sugiyama}
\affiliation{Neutron Science and Technology Center, 
Comprehensive Research Organization for Science and Society (CROSS), Tokai, Ibaraki 319-1106, Japan}
\author{Martin~M\aa nsson}
\affiliation{Department of Applied Physics, KTH Royal Institute of Technology, SE-106 91 Stockholm, Sweden}
\author{Yasmine~Sassa}
\email{yasmine.sassa@chalmers.se}
\affiliation{Department of Physics, Chalmers University of Technology, SE-41296 G\"oteborg, Sweden}

\date{\today}

\begin{abstract}
We report a muon spin rotation ($\mu^{+}$SR) study of the magnetic properties of the double perovskite compound LaSrNiReO$_{6}$. Using the unique length and time scales of the $\mu^{+}$SR technique, we successfully clarify the magnetic ground state of LaSrNiReO$_{6}$, which was previously deemed as a spin glass state. Instead, our $\mu^{+}$SR results point towards a long-range dynamically ordered ground state below $T_{\rm C}= 23$~K, for which a static limit is foreseen at $T=0$. Furthermore, between $23$~K~$<T\leq300$~K, three different magnetic phases are identified: a dense ($23$~K~$<T<75$~K), a dilute ($75$~K~$\leq T\leq250$~K), and a paramagnetic ($T>250$~K) state. Our results reveal how two separate, yet intertwined magnetic lattices interact within the unique double perovskite structure and the importance of using complementary experimental techniques to obtain a complete understanding of the microscopic magnetic properties of complex materials.
\end{abstract}


\maketitle

\section{\label{sec:Intro}Introduction}
Materials with perovskite crystal structure have for several decades been in the centre of attention across a wide scientific scope \cite{Jonker1956, Inaguma1994}. These compounds exhibit many interesting physical properties, such as various magnetic orders \cite{Wollan1955} and/or electronic states- metallic \cite{Warren1996, Torrance1992}, insulator \cite{Martinez1998}, and superconductivity \cite{Cava1988}. Moreover, some perovskites also display multiferroicity \cite{Wang2015, Valencia2011, Cheong2007, Sergienko2006}, which is an area that during recent years has received an increasing attention due to both fundamental interests as well as applications in sensors, actuators and memory devices \cite{Fiebig2016}.

The sm\"org\aa sbord of properties for the perovskites originate from its ABX$_3$-type crystal structure, where A and B are cations and X an anion that bond with B to form BX$_6$ octahedra. The significance of the perovskite structure is that the octahedra are flexible and can contract/expand/distort to accommodate almost all elements in the periodic table \cite{Woodward1997}. A noteworthy and the most common subgroup of perovskites is the oxide perovskite, which is achieved for systems where the X anion is an oxygen ion. 

Lately, a new type of perovskite has raised the interest of both experimental and theoretical physicists, the so-called double perovskite \cite{Kobayashi1998}. In this case, half of the B cations are substituted with another cation forming a A$_2$BB'O$_6$ structure. The B cation in these systems may order, where the most common pattern being a rock-salt type (like NaCl, but column or layered also exist), consisting of corner shared BO$_6$ and B'O$_6$ octahedra. \cite{ANDERSON1993, King2010} The degrees of freedom for designing perovskite compounds has opened up a wide door from application point of view. Lately, diverse combinations of compounds \cite{DEMAZEAU1993, YAMAMURA2006, Fresia1959} and physical properties for the double perovskite have been reported \cite{FU2000, Nakamura1971}. In such systems, the magnetic and electronic properties are governed by B and B' superexchange interaction through the O atom. Much like conventional perovskite systems, the double perovskites have been reported to exhibit metallic \cite{Tomioka2000}, insulator \cite{Paul2013}, superconductivity \cite{CHEN1997}, colossal magnetoresistance \cite{Kobayashi1998}, magnetic order \cite{Ritter2000}, frustrated magnetism \cite{Aharen2010} as well as multiferroicity \cite{Kumar2010}. 

\begin{figure*}[ht]
  \begin{center}
    \includegraphics[keepaspectratio=true,width=0.9\textwidth]{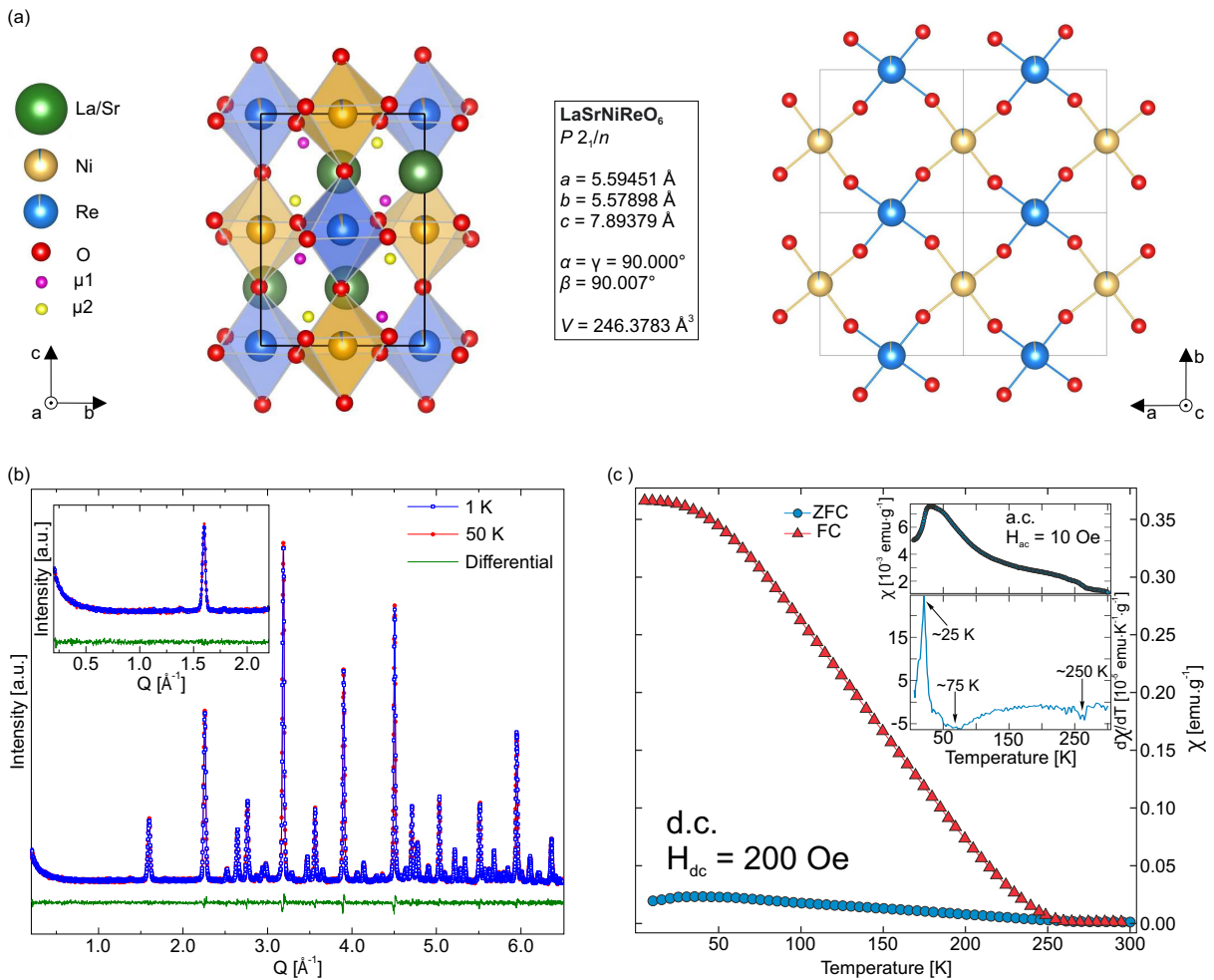}
  \end{center}
  \caption{
  Crystal structure, neutron powder diffraction pattern and magnetic susceptibility of LaSrNiReO$_6$. (a) On the left is the crystal structure of LaSrNiReO$_6$ with Ni-O and the Re-O octahedra displayed in orange and blue, respectively. The two predicted muon sites are included as magenta and yellow spheres: site $\mu1$ at $(0.21,0.26,0.875)$ and site $\mu2$ at $(0.26,0.21,0.625)$. On the right is a top view of the Re-O and Ni-O lattices. The Sr/La atoms are hidden for clarity. The Ni and Re atoms are displayed in orange and blue, while the Sr/La are in green and the oxygen in red. (b) Neutron powder diffraction data showing the patterns collected above (red filled circles) and below (blue open circles) the transition temperature. The inset is a zoom of the patterns between $Q$=~0.2 and 2.2~\AA$^{-1}$. The absence of magnetic Bragg peaks is made obvious with the difference plot displayed in green. (c) Zero field and field cooled (ZFC and FC) magnetic susceptibility of LaSrNiReO$_6$ measured under a field of $H= 200$~Oe. Inset shows the temperature dependence of the ZFC magnetic susceptibility in an applied field of 10~Oe, and its temperature derivative highlighting each transition.
  }
  \label{fig:Fig_1}
\end{figure*}

LaSrNiReO$_6$ is a double perovskite with two different magnetic ions, Ni (B) and Re (B') [see Figure \ref{fig:Fig_1} (a)]. In this particular system, an overlap between orbital symmetry is missing for an effective superexchange interaction. Instead, the ground state is determined by none or weakly interacting magnetic sub-lattices. A previous study by Ref. \onlinecite{Thompson2015} observed a frequency dependent shift in the cusp of a.c.-susceptibility, from where a spin glass ground state was suggested. Moreover, neutron diffraction studies indicated absence of long-range ordering \cite{Thompson2015, Somnath2018}, in line with a spin glass scenario. For the current sample, the absence of magnetic Bragg peaks in the neutron diffraction pattern are confirmed, as shown in Fig.~\ref{fig:Fig_1}(b). In order to further clarify the ground state, we initiated a muon spin rotation, relaxation and resonance ($\mu^{+}$SR) study. Being a local probe and highly sensitive to magnetism, $\mu^{+}$SR is the ideal tool to detect any weak magnetic interactions. Moreover, $\mu^{+}$SR allows for measurements in zero field (ZF) and/or weakly applied fields, meaning any influence from the measurements itself can be considered minimal, in comparison to $e.g.$ magnetic susceptibility measurements. Although, local perturbations may be induced, as in Pr-based materials (pyrochlores \cite{Foronda2015} or PrIn$_3$ \cite{Tashma1997}), where the muon causes an anisotropic local modification to the crystal field levels. In this study, a spin precession frequency is observed at the lowest temperature of $T= 2$~K, clearly excluding a spin glass scenario. Instead, an incommensurate long-range dynamically ordered ground state below $T_{\rm C}= 23$~K is proposed. Such state is still in line with previous neutron diffraction \cite{Thompson2015, Somnath2018} and $\chi^{AC}$ \cite{Thompson2015} results, as it will be shown. Furthermore, between 23~K and 300~K, our $\mu^{+}$SR results distinguish three other magnetic regimes, including dense, dilute, and paramagnetic states. Our findings demonstrate how two separate, yet intertwined magnetic lattices interact over a wide temperature range within the unique double perovskite structure. This study also establishes the unique capabilities of the $\mu^{+}$SR technique for investigating static and dynamic spins on the microscopic (local) length scale, and underlines the importance of combining complementary techniques to get insights on the true physical properties of complex materials. 

\section{\label{sec:exp} Experimental Methods}
A polycrystalline sample was prepared using a solid state reaction based on pure La$_2$O$_3$, SrCO$_3$, NiO, Re$_2$O$_7$ and Re metal as starting materials. Stoichiometric mixtures of the starting materials reacted in different steps at high temperatures, resulting in a single phased LaSrNiReO$_6$. Details about the synthesis and basic characterization of the sample are found in Ref. \onlinecite{Somnath2018}.

The crystal structure of LaSrNiReO$_6$ was generated using the Visualization for Electronic and STructural Analysis (VESTA) [\onlinecite{Vesta}] software. The magnetic susceptibility measurement was performed using both a Physical Property Measurement System (PPMS) and a SQUID magnetometer (MPMS) from Quantum Design. The d.c. and a.c. magnetic susceptibility were recorded as a function of temperature under a magnetic field $H_{\rm dc}= 200$~Oe and $H_{\rm ac}= 10$~Oe, respectively, within a temperature range of $T= 5-300$~K.

The neutron powder diffraction (NPD) experiments were performed at the High-Resolution Powder Diffractometer for Thermal Neutrons (HRPT) [\onlinecite{HRPT}] instrument at the Swiss Spallation Neutron Source (SINQ), Paul Scherrer Institute (PSI), Switzerland. About $1~$g of the sample was filled in a vanadium can and measured with two neutron wavelengths of $\lambda= 1.886$ and 2.95~\AA~at $T= 1$ and 50~K. 

The $\mu^{+}$SR experiments were performed at the surface muon beamlines M20 and DOLLY instruments at TRIUMF and PSI, respectively. Approximately, 1 g of sample was prepared inside a thin ($\sim50~\mu$m) aluminium coated mylar envelope, mounted on a Cu fork sample stick. A standard $^4$He flow cryostat was used in order to reach $T_{\rm base}=1.8$~K for PSI/DOLLY and $T_{\rm base}=2.4$~K for TRIUMF/M20. Finally, the software package \texttt{musrfit} was used in order to analyze the data \cite{musrfit}. 

\section{\label{sec:results}Results}
\subsection{Neutron Powder Diffraction and Magnetic Susceptibility}
Magnetisation and neutron powder diffraction measurements were performed [Fig. \ref{fig:Fig_1} (b-c)] before the $\mu^{+}$SR experiments. Starting with the $\chi$(T), two clear magnetic transitions are observed around $T= 250$~K and $T=25$~K, which agree with previous reports [\onlinecite{Thompson2015, Somnath2018}]. An additional transition around $T= 75$~K is made clear in the ZFC configuration and its temperature derivative (inset of Fig. \ref{fig:Fig_1} (c)). 
 
Figure \ref{fig:Fig_1} (b) displays the neutron diffraction patterns recorded at temperatures above and below the transition ($T\sim25$~K) and their difference plot. No magnetic Bragg peaks nor reducing of paramagnetic diffuse scattering are observed [see inset of Fig. \ref{fig:Fig_1} (b)] (only slight thermal expansion is seen) within this temperature range. It is worth mentioning that the paramagnetic diffuse scattering may not be observed within the experimental resolution of the present NPD measurement.  However, the ordered moments are sufficient to provide magnetic Bragg peaks (see Sec.~\ref{sec:discussion}), and their absences exclude a long-range magnetic order, which is in line with previous NPD measurements performed between $T= 2$ and $300$~K [\onlinecite{Thompson2015}, \onlinecite{Somnath2018}]. Detailed NPD and magnetisation analysis can be found in Ref. \onlinecite{Thompson2015, Somnath2018}. In order to clarify the nature of the ambiguous ground state, $\mu^{+}$SR measurements were performed at selected temperatures.  

\subsection{Muon Spin Rotation}
$\mu^{+}$SR measurements in zero field (ZF) and weak transverse field (wTF) configurations were performed. Here, the field in transverse directions refers to the applied field direction with respect to the initial $t_0$ muon spin polarisation, whereas the term weak signifies that the applied field is significantly weaker than the internal field at low temperatures. 

\subsubsection{Weak transverse field (wTF)}

\begin{figure}[ht]
  \begin{center}
    \includegraphics[keepaspectratio=true,width=75 mm]{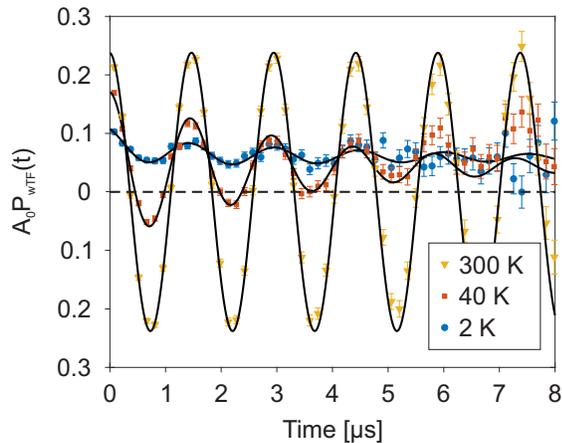}
  \end{center}
  \caption{
  Weak transverse field (wTF$=50$~G) time spectra recorded at $T$= 2~K, 40~K and 300~K. The solid lines are fits obtained using Eq. \ref{eq:wTF}. A clear oscillatory trends can be seen as function of temperature and the maximum asymmetry is fully recovered at 300~K.
  }
  \label{fig:wTFspec}
\end{figure}

The obtained time spectra for selected temperatures under ${\rm wTF} = 50$~Oe are presented in Fig. \ref{fig:wTFspec}. Apart from the oscillation resulting from the applied external magnetic field, an offset is observed at lower temperatures. Therefore, the wTF time spectra were fitted using an oscillatory component and two non-oscillatory depolarizing components according to
\begin{eqnarray}
 A_0 \, P_{\rm wTF}(t) &=&
A_{\rm TF} \cos(2\pi f_{\rm TF} t+\phi_{\rm TF})e^{-(\lambda_{\rm TF}t)} \cr &+& A_{\rm S}e^{(-\lambda_{\rm S} t)} + A_{\rm F}e^{(-\lambda_{\rm F} t)},
\label{eq:wTF}
\end{eqnarray}
where $A_{0}$ is the initial asymmetry and $P_{\rm TF}$ is the muon spin polarisation function in a wTF configuration. $A_{\rm TF}$, $f_{\rm TF}$, $\phi_{\rm TF}$ and $\lambda_{\rm TF}$ are the asymmetry, frequency, relative phase and depolarisation rate respectively, originating from the applied wTF. Further, $A_{\rm S}$, $\lambda_{\rm S}$, $A_{\rm F}$ and $\lambda_{\rm F}$ are the asymmetry and the respective depolarisation rates originating from internal magnetic fields. The indices S and F are conventions used to represent slow and fast components. 

\begin{figure}[ht]
  \begin{center}
    \includegraphics[keepaspectratio=true,width=73 mm]{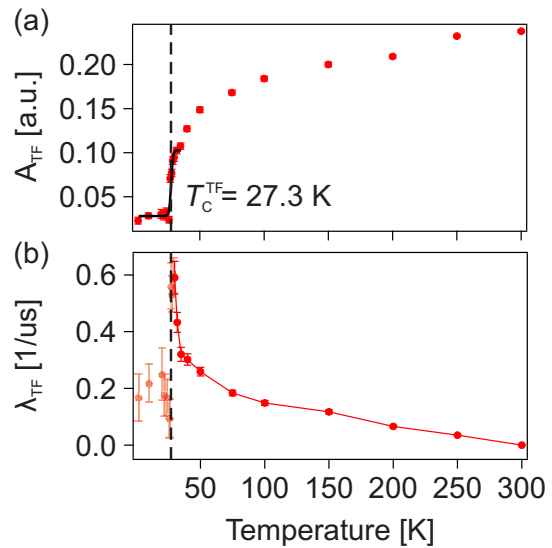}
  \end{center}
  \caption{
  Summary of the temperature dependent transverse field fitting parameters. (a) Transverse field asymmetry ($A_{\rm TF}$) versus temperature. The transition temperature at $T^{\rm TF}_{\rm C}= 27.3(3)$~K is deduced from a sigmoid fit (black solid line). The maximum asymmetry is only reached above $T= 250$~K. (b) Depolarization rate ($\lambda_{\rm TF}$) versus temperature. A clear increase of $\lambda_{\rm TF}$ is seen at the transition which decreases to near zero at high temperatures. The points below the transition (orange filled circles) are neglected [see text]. The solid line is a guide to the eye.  }
  \label{fig:wTF}
\end{figure}

The obtained fit parameters for the oscillatory components are summarized in Fig.~\ref{fig:wTF}. While the phase and the frequency remain constant through the whole temperature range, the amplitude of the oscillatory component ($A_{\rm TF}$) changes drastically. $A_{\rm TF}$ corresponds roughly to the paramagnetic fraction of the sample and the abrupt increase from low to high temperature represents a transition from a magnetically ordered to a disordered state. The transition temperature is then defined as the middle point of a sigmoid fit in which $T^{\rm TF}_{\rm C}=27.3(3)$~K is obtained. The full asymmetry is only recovered above 250~K, confirming a magnetic contribution to be present up to about 250~K as also suggested from the susceptibility data shown in Fig. \ref{fig:Fig_1}(c). The nature of this transition will be further discussed in Sec. \ref{sec:discussion}. As expected, the slow component ($A_{\rm S}$) is non-zero (not shown) until the full asymmetry of $A_{\rm TF}$ is recovered. Moreover, $A_{\rm TF}$ is low, but non-zero below the transition temperature ($A_{\rm TF}\simeq 0.02$ corresponding to $\sim$ 9\% of the signal), suggesting that part of the muon beam partially hits the sample holder and/or beamline, creating a small background contribution. 

The transverse field depolarisation rate, $\lambda_{\rm TF}$, approaches zero at higher temperatures as expected in the extreme motional narrowing limit of fluctuating magnetic moments. As the temperature is lowered, a critical behavior is displayed close to the transition with a sharp maximum around $T^{\rm TF}_{\rm C}$. Such behavior is consistent with critical slowing down of the magnetic moments and broadening of the internal field distribution. Given that the applied field is weak with respect to the internal field, the $\lambda_{\rm TF}$ values below $T^{\rm TF}_{\rm C}$ are not considered due to the demagnetisation field of the sample ($\lambda_{\rm TF}\simeq 0.2~\mu$s$^{-1}$ originates from the background signal).

\subsubsection{Zero field (ZF)}
\begin{figure}[ht]
  \begin{center}
    \includegraphics[keepaspectratio=true,width=75 mm]{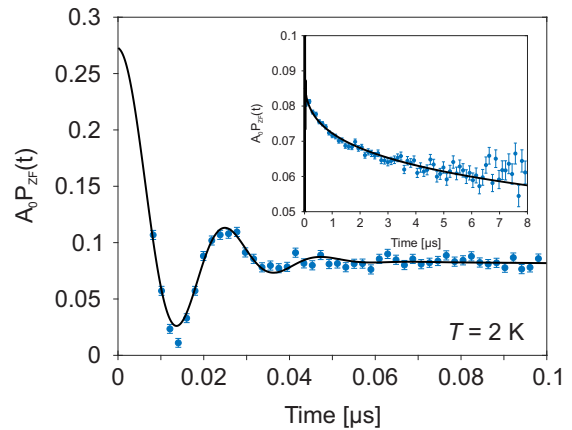}
  \end{center}
  \caption{Short time domain of the zero field (ZF) time spectrum recorded at $T_{\rm base}\simeq 2$~K, where a damped oscillation is seen. Inset shows the same data in the long time domain. The solid line is a fit of the ZF time spectrum using Eq. \ref{eq:ZF}.
  }
  \label{fig:ZFspec}
\end{figure}

\begin{figure}[ht]
  \begin{center}
    \includegraphics[keepaspectratio=true,width=75 mm]{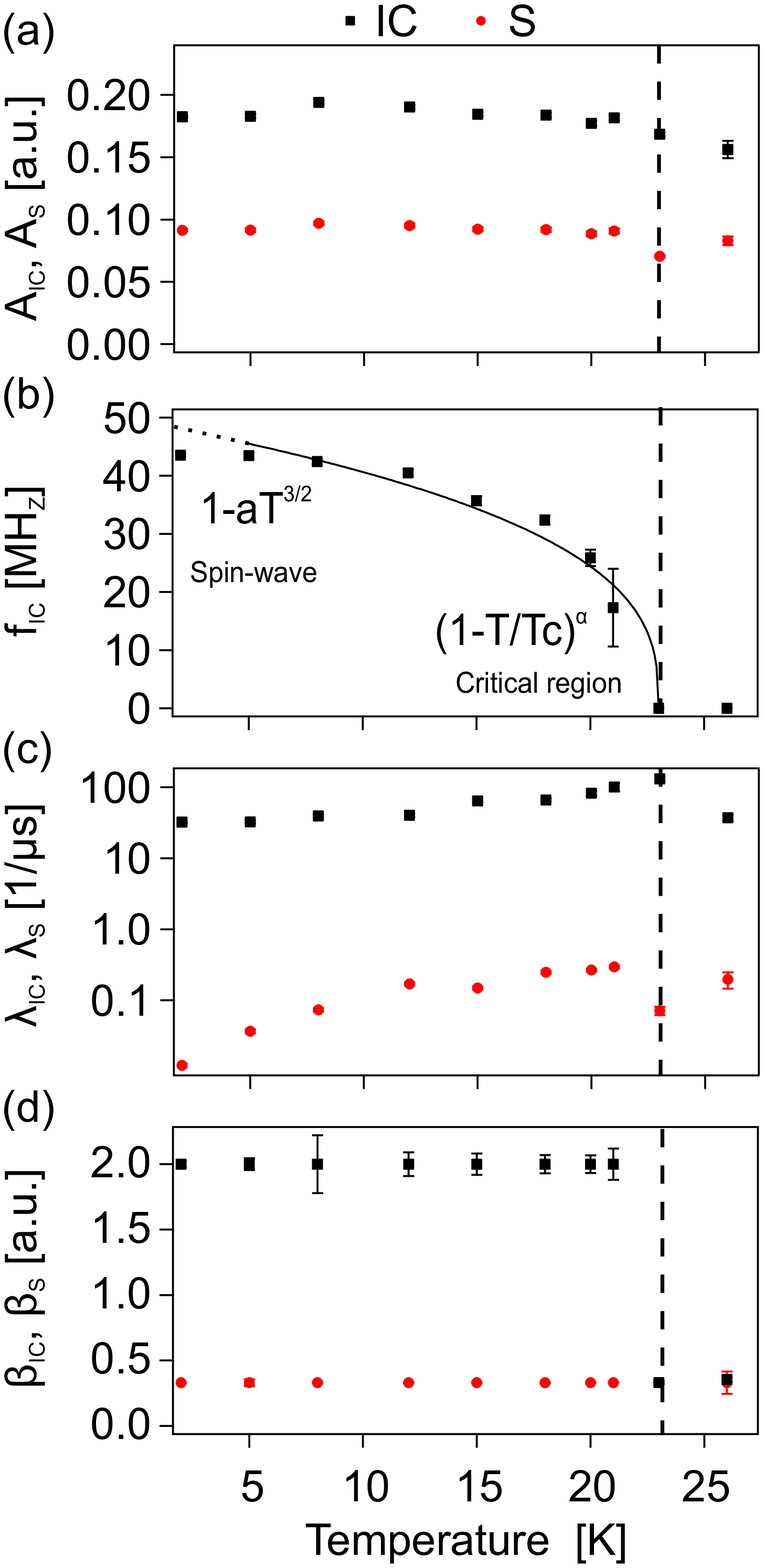}
  \end{center}
  \caption{
  Temperature dependence of the fit parameters from series of ZF time spectrum from $T= 2$~K to 27~K. (a) Asymmetries, (b) precession frequency, (c) depolarisation rates (log scale), and (d) stretched exponents. The solid line in (b) is a fit using $f({T})=f(0)(\frac{T_{\rm C}-T}{T_{\rm C}})^{\rm \alpha}$ down to $T=5$~K. The low temperature dotted part is extrapolated.
  }
  \label{fig:Slow}
\end{figure}

Figure \ref{fig:ZFspec} displays a ZF measurement at $T\simeq 2$~K. The shorter time domain exhibits a highly damped oscillation, which originates from which originates from field components perpendicular to the initial
muon polarisation. The longer time domain (inset Fig. \ref{fig:ZFspec}) reflects the spin dynamics (the so called tail component) and was best fitted by a stretched exponenetial. The time spectrum was then fitted using a combination of a stretched exponential and a Gaussian depolarising oscillating function
\begin{eqnarray}
 A_0 \, P_{\rm ZF}(t) &=&
A_{\rm IC} J_{0}(2\pi f_{\rm IC} t)e^{-(\lambda_{\rm IC}t)^{\beta_{\rm IC}}} + A_{\rm S}e^{-(\lambda_{\rm S} t)^{\beta_{\rm S}}}
\label{eq:ZF}
\end{eqnarray}

where A$_{0}$ is the initial asymmetry, P$_{\rm ZF}$ is the muon spin polarisation function in ZF configuration and A$_{\rm IC}$, $f_{\rm IC}$, $\beta_{\rm IC}$ and $\lambda_{\rm IC}$ are the asymmetry, frequency, stretched exponent and depolarisation rate for the oscillatory component, respectively. $J_{0}$ is the zero order Bessel function of its first kind, while A$_{\rm S}$, $\lambda_{\rm S}$ and $\beta_{\rm S}$ are the tail components originating from the fact that 1/3 of the field components inside the sample is parallel to the initial muon spin polarisation. The tail component exhibits a stretched exponential polarisation with a temperature dependent stretched exponent, $\beta_{\rm S}$. Physical interpretation of the stretched exponent is not trivial, but the function is derived by assuming a distribution of depolarization rates \cite{Steer2003, Johnston2006}. However, a microscopic origin for $\beta=1, 2, 0.5$ exists: $\beta=1$ denotes an exponential depolarization channel (‘magnetically homogenous’), while $\beta=2$ is a Gaussian depolarization channel (quasi-static field distribution) and $\beta=0.5$ is a so-called root exponential ($e.g.$ a distribution of spin correlation times, also seen above the transition temperature in spin glasses in the motional narrowing limit). A complete interpretation behind Eq.~\ref{eq:ZF} is further discussed in Sec. \ref{sec:discussion}.
\\~\\
\paragraph{Below the transition temperature $T=27$~K}
The ZF time spectra from $T\simeq 2$~K up to $T=27$~K were analysed using Eq.~\ref{eq:ZF} and the obtained fit parameters are displayed in Fig. \ref{fig:Slow}. At the base temperature, $A_{\rm S}\simeq A_0/3$ and $A_{\rm IC}\simeq2A_0/3$, consistent with being the tail and perpendicular components. It should be noted that the $t=0$ value of the asymmetry obtained is slightly overestimated because of lack of datapoints at low times. While the asymmetries [Fig. \ref{fig:Slow}(a)] are more or less constant up to the transition, drastic changes can be seen around the transition temperature for the other parameters. The temperature dependent muon precession frequency confirms an order parameter like dependence and displays a clear transition from an ordered to a disordered state [Fig. \ref{fig:Slow}(b)]. The value of the frequency is directly proportional to the internal field at the muon site. From mean field theory, a fit according to $f({T})=f(0)(\frac{T_{\rm C}-T}{T_{\rm C}})^{\rm \alpha}$ results in $T^{\rm ZF}_{\rm C}=23.0(1)$~K, $\alpha=0.348 (47)$ and $f(0)=49.57(2.49)$. The low temperature part is not well fitted due to thermal magnon excitation, resulting in a reduction of local spontaneous magnetization given by the Bloch 3/2 law \cite{Blundell2003}. Note that the slightly higher transition temperature extracted from the wTF measurement ($T^{\rm TF}_{\rm C} = 27.3$~K) may be due to the applied field and in general, only the ZF data reflect the intrinsic magnetic properties of materials.

Both depolarisation rates [Fig. \ref{fig:Slow}(c) plotted for the ordinate axis in log scale] seem to have similar temperature dependence, where the values are increasing as $T$ increases, consistent with an increase of dynamics close to the phase transition. The parallel component, $\lambda_{\rm S}$, corresponds roughly to the spin-lattice relaxation rate and contains information about the dynamics in the system, while $\lambda_{\rm IC}$ includes a mixture of both the field distribution and the dynamics. Therefore, the relatively high value of $\lambda_{\rm IC}=32.4(1.2)$~$\mu$s$^{-1}$ at $T_{\rm base}$ relates to a high field distribution and dynamics at the muon sites. However, the behavior of $\lambda_{\rm S}\rightarrow 0$ as T$\rightarrow 0$, is consistent with the sample's magnetism going towards the static limit. Since $\lambda_{\rm IC}$ seems to level off at lower temperatures, the high value of $\lambda_{\rm IC}$ originates mostly from a broad field distribution. Based on the temperature dependence of $\lambda_{\rm S}$, a completely static magnetic ground state is expected below $T=2$~K. Nevertheless, the value $\lambda_{\rm S}(2\rm$~K$)=0.012(1)$~$\mu$s$^{-1}$ is still observed, meaning that the spins are dynamic even at lowest measured temperature. Finally, the stretched exponents are almost constant up to the transition. The stretched exponent of $\beta=2$ is a Gaussian depolarization channel suggesting that the magnetic phase is quasi-static. $\beta = \frac{1}{3}$ corresponds to the magnetic impurity limit and has already been reported in several spin glass systems \cite{Amit1996, Campbell1994, Ogielski1985}.  
\\~\\
\paragraph{Above the transition temperature $T=27$~K}
\begin{figure}[ht]
  \begin{center}
    \includegraphics[keepaspectratio=true,width=75 mm]{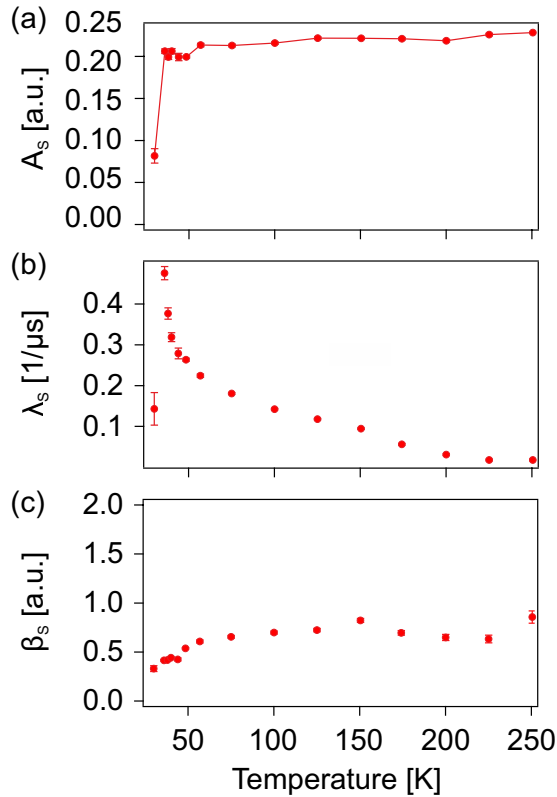}
     \end{center}
  \caption{Temperature dependence of the fit parameters from series of ZF time spectrum from $T= 27$ to 250~K: (a) Asymmetry, (b) depolarisation rate and (c) stretched exponent. The solid line in (a) is a guide to the eye. 
  }
  \label{fig:HTSlow}
\end{figure}

\begin{figure}[ht]
  \begin{center}
    \includegraphics[keepaspectratio=true,width=75 mm]{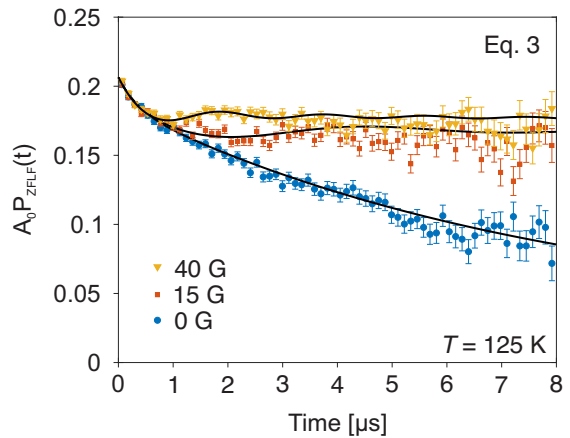}
     \end{center}
  \caption{Zero field (ZF) and longitudinal field (LF$ = 15$ and $20$~G) time spectra with best fits using Eq.~\ref{eq:ZFKT} are shown as solid black lines.
  }
  \label{fig:ZFLFSpec}
\end{figure}

The sample was also studied in ZF configuration for temperatures above $T^{\rm ZF}_{\rm C}$ up to $T=250$~K. The data are also well fitted using Eq.~\ref{eq:ZF} in this temperature range (note that $A_{\rm IC} =0 $ above $T^{\rm TF}_{\rm C}$). The obtained fit parameters as a function of temperature are shown in Fig. \ref{fig:HTSlow}. A clear increase in asymmetry is shown with increasing temperature until the maximum asymmetry $A_0\simeq0.24$ is recovered. The depolarization rate is exhibiting a critical behavior just above $T^{\rm ZF}_{\rm C}$. The narrow temperature range of this critical slowdown of electronic moments suggests an exchange coupling $J\sim k_{\rm B}\rm T_{\rm C}$. As for the stretched exponent, a recovery of $\beta \rightarrow 1$ is observed at the highest temperature, as expected. Moreover, $\beta \rightarrow \frac{1}{3}$ is seen around the transition, typical for many glassy like transitions \cite{Amit1996, Campbell1994, Ogielski1985}. Longitudinal field measurements (LF$ = 15$ and 40~G) show that  Eq.~\ref{eq:ZF} may not suitable in the temperature range of $75$~K$-250$~K where $\beta \sim 0.7$, because the value of $\beta$ changes dramatically with the applied field (see Apendix~\ref{Appendix} and Fig.~\ref{fig:ZFLFSpec_apen}). Instead, the ZF+LF time spectra were nicely fitted (Fig.~\ref{fig:ZFLFSpec}) with the following function for this temperature range

\begin{eqnarray}
 A_0 \, P_{\rm ZF}(t) &=&
A_{\rm F} e^{-\lambda_{\rm F}t} + A_{\rm KT}L^{SLKT}(\Delta, t),
\label{eq:ZFKT}
\end{eqnarray}

where $A_{0}$ is the initial asymmetry, $P_{\rm ZF}$ is the muon spin polarisation function in ZF configuration, $A_{\rm F}$ and $A_{\rm KT}$ are the asymmetries for their respective contribution, where L$^{SLKT}$ represents a static Lorentzian Kubo-Toyabe. $\lambda_{\rm F}$ is the depolarisation rate for an initial fast depolarising signal, while $\Delta$ is related to the internal Lorentzian half width half maximum. The L-KT is the dilute limit of the KT, commonly observed in paramagnets with pressence of dillute magnetic moments \cite{Walstedt1974, Lamura2014}. $A_{\rm F}$ is therefore attributed to the fraction of muons sitting close to these dilute magnetic highly fluctuating moments. Indeed, it is only for $T>250$~K that a Gaussian KT could fit the data, meaning that dilute electronic magnetic moments are present up to this temperature. Furthermore, the data below $75$~K could not be fitted to any type of Kubo-Toyabe functions (i.e. dynamic/static Gaussian or Lorenzian distributed KT). This suggests the presence of a distribution of relaxation times for $23$~K$<T<75$~K. Eq.~\ref{eq:ZFKT} is further discussed in Sec.~\ref{sec:discussion}.

\begin{figure}[ht]
  \begin{center}
    \includegraphics[keepaspectratio=true,width=75 mm]{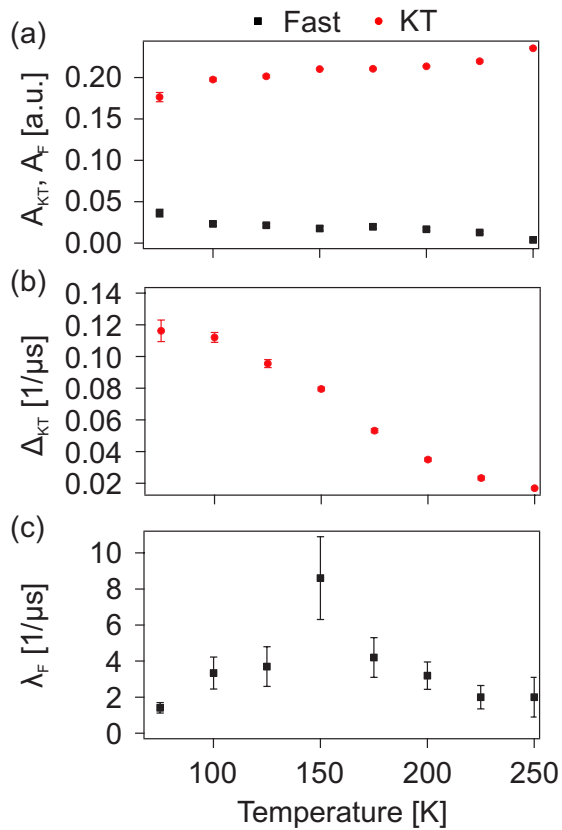}
     \end{center}
  \caption{
  Temperature dependence of the fast and Kubo-Toyabe (KT) fit parameters from Eq.~\ref{eq:ZFKT}, obtained from series of ZF time spectrum from $T= 70$~K to 250~K: (a) Asymmetry, (b) field distribution width and (c) depolarization rate.
  }
  \label{fig:HTKT}
\end{figure}

The obtained fit parameters using Eq. \ref{eq:ZFKT} are displayed in Fig. \ref{fig:HTKT}. The KT asymmetry slowly recovers the full asymmetry as the temperature increases, while the fast relaxing component decreases, suggesting that number of dillute moments are decreasing with temperature. Naturally, this decrease will also show in the temperature dependence of $\Delta$. In principle, Eq.~\ref{eq:ZFKT} may be replaced be replaced by the dynamical L-KT. Although, dynamical contribution are usually observed in the tail of the KT, which is not accessible in the present data. Instead, the dynamical contribution is affecting the field distribution width, which is decreasing with increasing temperature. Strangely, the depolarisation rate ($\lambda_{\rm F}$) seems to exhibit a maximum around 150~K. Indeed, a small anomaly can also be observed around this temperature in the stretched exponent, shown in Fig. \ref{fig:HTSlow}(c). The origin of this anomaly is currently unknown and further investigations are required. Although, such anomaly does not affect the main results and conclusion drawn in this report. 

\begin{figure*}[ht]
  \begin{center}
    \includegraphics[keepaspectratio=true,width=\textwidth]{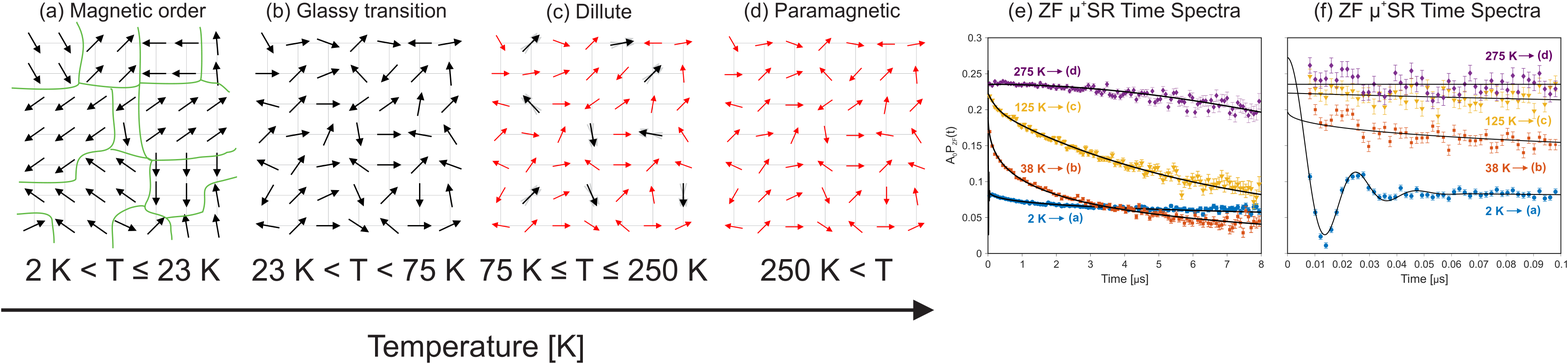}
  \end{center}
  \caption{The different magnetic phases and the corresponding ZF $\mu^+$SR time spectra highlighted in this study with schematic representation of electronic moments (black) and nuclear moments (red). (a) Magnetic order with a distinct correlation length [see text], (b) dense or glassy like transition (c) dilute state, (d) paramagnetic phase and ZF time spectra at respective temperatures for each highlighted phase in (a-d) in short (e) and long (f) time domains. The solid black lines represents with with Eq.~\ref{eq:ZF} (2~K and 38~K), Eq.~\ref{eq:ZFKT} (125~K) and $A_{\rm KT}G^{\rm SGKT}(\Delta,t)$ (275~K) where $G^{\rm SGKT}$ represents a static Gaussian Kubo-Toyabe.
  }
  \label{fig:Schematic}
\end{figure*}

\section{\label{sec:discussion}Discussion}
Based on the results presented in Sec. \ref{sec:results}, the following four magnetic states are identified (Fig. \ref{fig:Schematic}): (1) a magnetically ordered state below $T^{\rm ZF}_{\rm C}=23$~K, (2) a glassy like transition to a random dense magnetic state between 23~K~$<T<75$~K, (3) a random dilute magnetic phase between 75~K~$\leq T\leq$~250~K, (4) a paramagnetic state above $T>$250~K. The presence of several magnetic phases is further supported by the magnetic susceptibility measurements shown in Fig.~\ref{fig:Fig_1} (c) and Ref.~\onlinecite{Somnath2018}, where both the transition at $T\approx 23$~K and the bifurcation between ZFC and FC curves at $T\approx 250$~K were reported. The latter has also been observed in Sr$_2$CaReO$_6$ \cite{Wiebe2002} and Sr$_2$InReO$_6$ \cite{Gao2011} double perovskite compounds and attributed to the non-magnetic ions located at the B-site, causing a geometrical frustration of the Re site on the FCC sub-lattice. Such situation would lead to a dilute magnetic system and a L-KT like depolarisation would manifest the ZF time spectrum in $\mu^+$SR, as presented here. Typically, a L-KT fit is appropriate for dilute electronic moments existing in a non-magnetic matrix. Moreover, the weak temperature dependence of $\lambda_{\rm S}$ from ZF measurement supports such situation in which dilute moments fluctuate [Fig.~\ref{fig:HTSlow}(b)]. 

In the same temperature region, a stretched exponent of $\beta \sim 0.7$ was obtained. While the stretched exponential is somewhat phenomenological, such value is perhaps indicative of an intermediate case between dilute and dense motionally narrowed source of magnetic fields. If the magnetic distribution in real space is dense, the field distribution can be approximated by a Gaussian shape for which the muon spin depolarization follows an exponential in the motional narrowing limit. In the dilute limit, the muon spin depolarizes according to a root exponential in the narrowing limit. Therefore, a $\beta \sim 0.7$ for 75~K$\leq T\leq$250~K could perhaps be explained as an intermediate case of dense and dilute source of magnetic fields. \cite{Campbell1994, Noakes2002}

A stretched exponential like depolarization is observed in a wide temperature range ($23$~K~$<T<75$~K). Such a situation is usually explained by the presence of a distribution of muon depolarization channels (i.e. spatially disordered systems). The magnetically similar compounds Sr$_2$NiWO$_6$ and Sr$_2$NiTeO$_6$ possess non-magnetic W$^{6+}$ and Te$^{6+}$ ions at $B'$ site. The Ni 3$d$-moments in these compounds orders at $T_{\rm N}=35$~K and $T_{\rm N}=54$~K, respectively \cite{Iwanaga2000}. Therefore, it seems that some localized moments of Ni$^{2+}$ becomes prominent at lower temperature for the title compound, effectively destroying the dilute limit and resulting into a distribution of relaxation rates. Therefore, we suggest that an independent Ni sublattice feature is realized below $T^{\rm ZF}_{\rm C}=23$~K, resulting into magnetic order. In other words, the Re$^{5+}$ interactions/fluctuations dominate at higher temperature, while the Ni$^{2+}$ interactions become significant at lower temperature. Admittedly, the presented data cannot distinguish the Ni$^{2+}$ moments from Re$^{5+}$ moments, meaning that the opposite case in which the Re$^{5+}$ orders at low $T$ instead of Ni$^{2+}$ is also probable. 

Transport measurements \cite{Somnath2018} suggested that the transition at $T\sim30$~K is due to a weak ferromagnetic interaction, predicted by Goodenough-Kanamori rules. However, in this study, a Bessel function was used instead of a simple cosine function to fit the time spectrum. This would suggest an incommensurate magnetic ordering at low temperatures. This is justified because using a cosine function results in a large offset in the initial phase $\phi \approx-50^{\circ}$ (see Appendix~\ref{Appendix}). For a particular case of an incommensurate single-k collinear magnetic structure the polarisation function is given by:

\begin{eqnarray}
J_0(\gamma_\mu B_{\max}t)\simeq \sqrt{\frac{2}{\pi\gamma_\mu B_{\rm max}t}}cos(\gamma_\mu B_{\rm max}t-\pi/4)
\label{eq:Bessel}
\end{eqnarray}

A phase offset may also be realized in magnetic structures, $e.g.$ in cases where the local field fluctuates from parallel to anti-parallel direction at a rate $v_{\rm c} < 2 \gamma_\mu B_{\rm fluc}$ \cite{Yaouanc2011}. However, the spin-lattice relaxation rate [$\lambda_{\rm S}$ in Fig.~\ref{fig:Slow}(c)] indicates a decrease in dynamics with temperature, which in turn would change the phase offset as a function of temperature (not this case). Another case was reported by Ref. \onlinecite{Sugiyama2009}, where several muon sites were present within a larger magnetic unit cell. While the presence of several muon sites could explain the stretched exponential behavior in this study, only one clear oscillating signal was observed meaning only one (or possibly two magnetically similar) muon site is expected for this compound. Consequently, LaSrNiReO$_6$ most likely display an incommensurate magnetic spin order below $T^{\rm ZF}_{\rm C}$. 

It should also be noted that the muon spin precession frequency at the lowest temperature was effectively lowered due to the reduction in spontaneous magnetisation (Fig.~\ref{fig:Slow}(b)) \cite{Blundell2003}. However, such situation is only realised for ferro and ferri magnets. Indeed, the transition at $T\sim30$~K was predicted to be due to a weak ferromagnetic interaction \cite{Somnath2018} according to Goodenough-Kanamori rules. Consequently, given the fact that we observe an incommensurate order, the ground state of LaSrNiReO$_{6}$ points towards an  incommensurate ferrimagnetic state. 

The presence of an oscillatory component in the $\mu^+$SR time spectrum for $T\leq23$~K (Fig. \ref{fig:ZFspec}) suggests a magnetically ordered ground state, where $\chi^{\rm AC}$ measurements also predicted a magnetic transition occurring around 27~K. However, magnetic Bragg peaks were not observed in the neutron diffraction measurements \cite{Thompson2015, Somnath2018} and such discrepancy should be addressed. First off, let us point out that the internal field distribution is wide, as implied by the highly damped oscillation ($\lambda_{\rm IC}$). As mentioned, the parallel component of the ZF measurement, $\lambda_{\rm S}$, indicated a dynamic state at the base temperature. While it is possible for spins to be dynamic in only specific directions, it is noted that the measurement was performed on a powder sample meaning any spatial direction should be averaged out. Furthermore, $\lambda_{\rm IC}$ seems to level off at lower temperatures while $\lambda_{\rm S}$ continues to decrease [Fig.~\ref{fig:HTSlow} (c)]. Therefore, the high damping of the oscillatory component, $\lambda_{\rm IC}$, observed in Fig. \ref{fig:Slow} should mainly be due to a wide field distribution width and supports the fact that the sample is magnetically inhomogeneous. Naturally, magnetically in-homogeneous systems do not yield any clear magnetic Bragg peaks in NPD, which would suggest a short-range magnetic order. However, the observation of an oscillation in ZF $\mu^+$SR time spectrum points towards a long-range magnetic order.

The absence of magnetic Bragg peaks in the low-temperature NPD pattern could be connected to a weak ordered moment. The ordered moment can be estimated from the present $\mu^+$SR data and by predicting the possible muon site using the density functional theory (DFT) package $Quantum~espresso$ \cite{QE-2009, QE-2017}. For such calculation, (i) the chemical disorder was not considered, (ii) any local perturbations due to the muon were not considered and (iii) the muon site was assumed to be at the electrostatic potential minimum. Given the above assumption, a self consistent calculation using the psudepoentials described by Ref. \cite{Lejaeghere2016, Prandini2018} provides two possible moun site candidates $\mu1$ at $(0.21,0.26,0.875)$ and $\mu2$ at $(0.26,0.21,0.625)$ [see Fig.~\ref{fig:Fig_1}]. Assuming now that the magnetic structure is the same as that of the sister compound LaCaNiReO$_6$ \cite{Somnath2018} and that the local field is solely composed of dipolar fields ($B_{\rm dip}$), the expected internal field at these sites are calculated to be $f_1=45.31$~MHz and $f_2=42.49$~MHz ($B_{\rm dip}=f\gamma_{\mu}$). For these internal fields, the Ni$^{2+}$ and Re$^{5+}$ moments were set to $\mu_{\rm Ni}=2~\mu_{\rm B}$ and $\mu_{\rm Re}=1~\mu_{\rm B}$, respectively. Indeed, the expected internal field for the assumed moment sizes is very close to the experimental one $f(0~$K$)=49.57(2.49)$~MHz (Fig.~\ref{fig:Slow}). Therefore, we deduct that the ordered moment is strong enough to be detected by the NPD experiment.

A previous report based on a.c.-susceptibility measurements suggested a spin glass state below $T\sim 25$~K \cite{Thompson2015}. Indeed, just like in this case, many spin glasses exhibit $\beta\rightarrow\frac{1}{3}$ around the transition temperature. However, presence of a muon spin precession clearly excludes such scenario. Instead, we suggest that an incommensurate order is stabilised with a distinct magnetic correlation length ($\xi$), that is long enough for detection for $\mu^+$SR and $\chi^{\rm AC}$ techniques, but not for neutron diffraction (hence magnetically in-homogeneous). On a macroscopic scale, the system may look like a spin glass, where small magnetic domains freeze randomly. However, microscopically, each of these domains are in fact ordered on a shorter length scale [Fig.~\ref{fig:Schematic}(a)]. Overall, this kind of microscopic picture would yield a frequency dependent shift in $\chi^{\rm AC}$ while also yielding muon spin precession and the absence of magnetic Bragg peaks in neutron diffraction is explained. From our knowledge, this is a very rare case where such a situation is present and so clearly revealed.

\section{\label{sec:conclusion}Conclusions}
We have utilised muon spin rotation ($\mu^{+}$SR) to elucidate the magnetic properties of the double perovskite compound LaSrNiReO$_{6}$. Using the unique length and time scales of the $\mu^{+}$SR technique, we have successfully identified four magnetic states: a paramagnetic ($T>250$~K), a dilute ($75$~K~$\leq T\leq250$~K), a dense ($23$~K~$<T<75$~K), and an incommensurate magnetically ordered state ($2$~K~$<T\leq23$~K). The dilute state is established by weakly interacting and fluctuating Re$^{5+}$ ions sitting on the $B'$ site, which develops into an incommensurate order around 23~K, driven by the FCC sublattice of Ni ions on the B sites. This state consist of weakly interacting domains/islands, established by a ferri interacting spins, forming an incommensurate spin wave. This study reveals in great detail how two separate, yet intertwined magnetic lattices interact over a wide temperature range within the unique double perovskite structure. It also shows the unique capabilities of the $\mu^{+}$SR technique for studying static and dynamic spins on the microscopic (local) length scale. We also emphasize the importance of applying a set of complementary experimental techniques in order to obtain a complete and correct understanding of the microscopic magnetic properties in complex materials.

\begin{acknowledgments}
This research was supported by the European Commission through a Marie Sk{\l}odowska-Curie Action and the Swedish Research Council - VR (Dnr. 2014-6426 and 2016-06955) as well as the Carl Tryggers Foundation for Scientific Research (CTS-18:272). J.S. acknowledge support from Japan Society for the Promotion Science (JSPS) KAKENHI Grant No. JP18H01863. Y.S. is funded by the Swedish Research Council (VR) through a Starting Grant (Dnr. 2017-05078) and E.N. the Swedish Foundation for Strategic Research (SSF) within the Swedish national graduate school in neutron scattering (SwedNess). Y.S. and K.P. acknowledge funding a funding from the Area of Advance- Material Sciences from Chalmers University of Technology. D.A. acknowledges partial financial support from the Romanian UEFISCDI Project No. PN-III-P4-ID-PCCF-2016-0112. The neutron diffraction experiments
were performed at the Swiss spallation neutron source (SINQ)
at the Paul Scherrer Institute, Villigen, Switzerland. We thank P. Anil Kumar for the contribution to this work and Andreas Sutter for explaning how the error bars are calculated in $musrFIT$. Finally, we greatly acknowledge the very valuable discussions with Prof. Jess Brewer concerning the $\mu^{+}$SR fitting functions and procedures and Prof. Alexei Kalaboukhov for measurement time at the PPMS. All images involving crystal structure were made with the VESTA software \cite{Vesta}
\end{acknowledgments}

\appendix
\section{Appendix: ZF fitting procedure at low temperatures}
\label{Appendix}
\begin{figure}[ht]
  \begin{center}
    \includegraphics[keepaspectratio=true,width=75 mm]{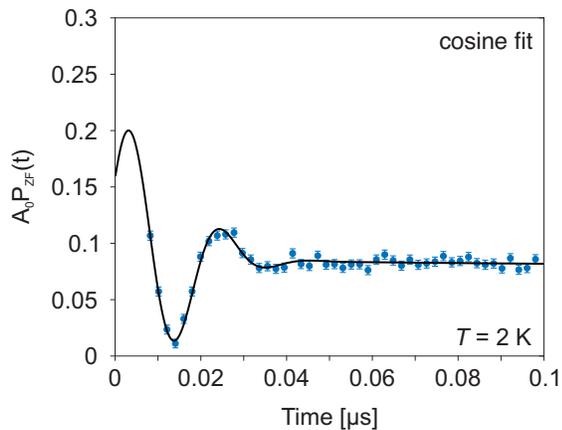}
  \end{center}
  \caption{Short time domain of the zero field (ZF) time spectrum recorded at $T_{\rm base}\simeq 2$~K, where a damped oscillation is seen. The solid line is a fit of the ZF time spectrum using $A_0 P_{\rm ZF}(t) =
A_{\rm C} cos(2\pi f_{\rm C} t+\phi)e^{-(\lambda_{\rm C}t)^{2}} + A_{\rm S}e^{-(\lambda_{\rm S} t)^{\beta_{\rm S}}}$: $A_{\rm C}=0.115(7)$, $\lambda_{\rm C}=48(2)~\mu$s$^{-1}$, $\phi=-54(2)^{\circ}$, $f_{\rm C}=45(1)$~MHz, $A_{\rm S}=0912(4)$, $\lambda_{\rm S}=0.0121(7)~\mu$s$^{-1}$ and $\beta_{\rm S}=0.330(3)$
  }
  \label{fig:ZFspec_cos}
\end{figure}

Equation~\ref{eq:ZF} is composed of a zero order Bessel function of its first kind, instead of a simple cosine function. This was justified because a cosine function yields unreasonable large values of phase. For clarity, a fit using a cosine function is shown in Fig.~\ref{fig:ZFspec_cos} and a clear phase shift is observed. As a result, a zero order Bessel function of its first kind was chosen instead of a simple cosine function. Additionally, the asymmetry of the oscillatory component is underestimated with respect to $A_0$ for the cosince function. Although, a small overestimation is observed with fits using Eq.~\ref{eq:ZF}. The choice of oscillatory component does not affect the tail component.

\begin{figure}[ht]
  \begin{center}
    \includegraphics[keepaspectratio=true,width=75 mm]{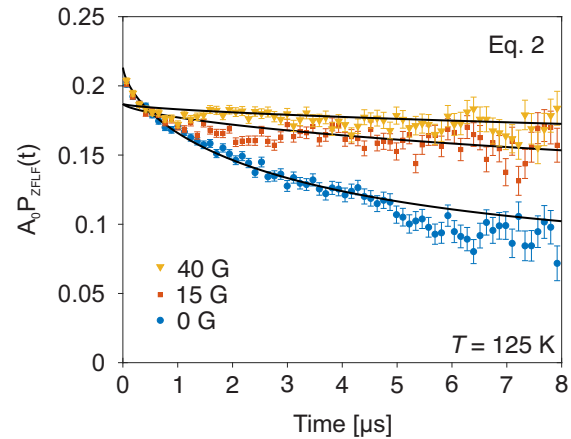}
     \end{center}
  \caption{Zero field (ZF) and longitudinal field (LF$ = 15$ and $20$~G) time spectra with best fits using Eq.~\ref{eq:ZF} in (a) are shown as solid black lines. A common $\beta$ was utilized for all field configurations since the spin-spin correlation should be unaffected by such weak LFs.
  }
  \label{fig:ZFLFSpec_apen}
\end{figure}

Figure~\ref{fig:ZFLFSpec_apen} show the ZF+LF time spectrum collected at $T=125$~K. As decribed in the main text, a fit using Eq.~\ref{eq:ZF} results in a wide changes in the parameter $\beta$: 0.66(3) at ZF, 0.143(8) at LF$=15$~G and 0.102(7) at LF$=40$~G. The obtained values under LF is unreasonably small and more importantly, a large change is seen between ZF and LF suggesting that the spin-spin correlation would change with LF$=15$~G. If the parameter is instead shared among the field configurations, a $\beta=0.65(3)$ is obtained instead. However, the quality of fit is greatly reduced, as seen in Fig.~\ref{fig:ZFLFSpec_apen}. As a result, Eq.~\ref{eq:ZFKT} was selected for fitting the ZF time spectra in the temperature range $T= 75 - 250$~K.

\bibliography{Refs} 
\end{document}